\documentclass[12pt]{article}

\usepackage{fancybox}

\usepackage{cite}
\usepackage{float}
\usepackage{amsfonts}
\usepackage{amsmath}
\usepackage{amsbsy}
\usepackage{graphicx}
\usepackage{amssymb}
\usepackage{amsthm}
\usepackage{bm}
\usepackage{epsfig}
\usepackage{latexsym}
\usepackage{pdflscape}
\usepackage{color}
\numberwithin{equation}{section}

\allowdisplaybreaks

\DeclareMathOperator{\tr}{tr}
\DeclareMathOperator{\pf}{pf}
\DeclareMathOperator{\Ai}{Ai}
\DeclareMathOperator{\vol}{vol}

\newcounter{aff}

\begin{document}

\begin{titlepage}
\begin{flushright}
{\footnotesize OCU-PHYS 425, YITP-15-18}
\end{flushright}
\begin{center}
{\Large\bf Superconformal Chern-Simons Partition Functions\\[6pt]
of Affine D-type Quiver from Fermi Gas}

\bigskip\bigskip
{\large 
Sanefumi Moriyama\footnote[1]{{\tt moriyama@sci.osaka-cu.ac.jp}\\
Address before March 2015: Kobayashi Maskawa Institute
\& Graduate School of Mathematics,\\
Nagoya University, Nagoya 464-8602, Japan}
\quad and \quad
Tomoki Nosaka\footnote[2]{\tt nosaka@yukawa.kyoto-u.ac.jp}
}\\
\bigskip
${}^{*}$\,
{\small\it Department of Physics, Graduate School of Science,
Osaka City University,\\
Osaka 558-8585, Japan}
\medskip\\
${}^{\dagger}$\,
{\small\it Yukawa Institute for Theoretical Physics,
Kyoto University,\\
Kyoto 606-8502, Japan}
\end{center}

\begin{abstract}
We consider the partition function of the superconformal Chern-Simons theories with the quiver diagram being the affine D-type Dynkin diagram.
Rewriting the partition function into that of a Fermi gas system, we show that the perturbative expansions in $1/N$ are summed up to an Airy function, as in the ABJM theory or more generally the theories of the affine A-type quiver.
As a corollary, this provides a proof for the previous proposal in the large $N$ limit.
For special values of the Chern-Simons levels, we further identify three species of the membrane instantons and also conjecture an exact expression of the overall constant, which corresponds to the constant map in the topological string theory.
\end{abstract}

\bigskip\bigskip\bigskip

\centering
\includegraphics[width=10cm]{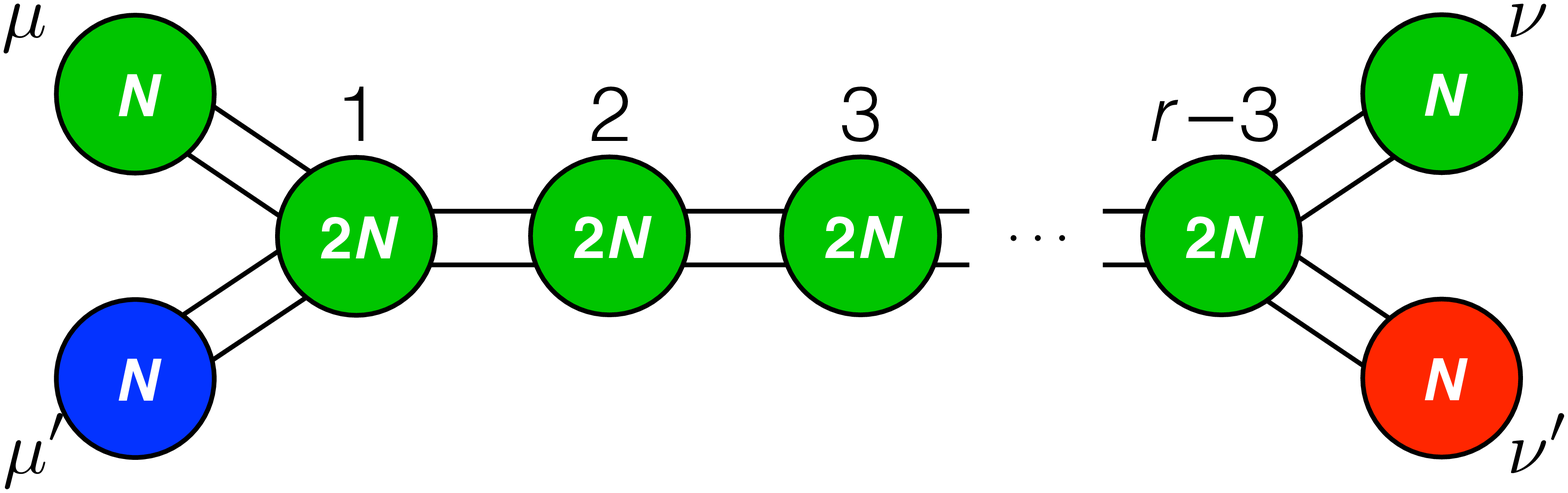}

\end{titlepage}
\tableofcontents

\section{Introduction}\label{introduction}

The $ADE$ classification not only provides a beautiful structure in the mathematical science, but also plays important roles in M-theory.
As an example related to the M5-brane, there is an $ADE$ classification of six dimensional ${\cal N}=(2,0)$ theories, coming from the classification of the singularities on which the M5-branes are placed \cite{W2}.
Also, even for the six dimensional ${\cal N}=(1,0)$ theories, the $ADE$ classification continues to be crucial \cite{HMRV}.

In the context of the M2-branes, the three dimensional $U(N)$ quiver gauge theories can also be classified by the affine $ADE$-type, or ${\widehat A}{\widehat D}{\widehat E}$-type, Dynkin diagrams.
A large class of three dimensional superconformal Chern-Simons theories can be constructed by quiver diagrams as follows.
For each vertex in the quiver diagram, we assign a vector multiplet of a gauge group $U(N)$, with $N$ proportional to the comark.
For each edge connecting two vertices, we assign a pair of hypermultiplets which are in the bifundamental representation under the gauge groups on the two vertices.
For example, the simplest quiver, ${\widehat A}_1$, gives the ABJM theory \cite{ABJM}.
For these theories, the localization technique allows us to express the partition function on $S^3$ as a finite dimensional matrix model \cite{KWY}.
Then, it was shown in \cite{GAH} that, if we require that the long range force among the eigenvalues vanishes (also known as the balance condition in \cite{GW}), the quivers have to be of ${\widehat A}{\widehat D}{\widehat E}$-type.
The behaviour of these matrix models in the limit $N\rightarrow\infty$ was studied for the ${\widehat A}$-type quivers in \cite{HKPT,GHP}, for the ${\widehat D}$-type quivers in \cite{GAH,GHN,CHJ} and for the ${\widehat E}$-type quivers in \cite{CHJ}.
Interestingly, they observed a universal scaling law in the partition function for all the ${\widehat A}{\widehat D}{\widehat E}$-type quivers
\begin{align}
Z(N)\sim\exp\biggl[{-\frac{2}{3\sqrt{C}}N^{\frac{3}{2}}}\biggr],
\end{align}
with $C$ some constant depending on the quivers.
This scaling law is a characteristic property of the M2-branes in the context of the AdS/CFT correspondence \cite{KT}.

For the ABJM theory, after this leading $N^{\frac{3}{2}}$ scaling behaviour was obtained in \cite{DMP1}, the partition function was studied in full detail.
It was shown that the $N^{\frac{3}{2}}$ behaviour is completed by the perturbative $1/N$ corrections into an Airy function \cite{FHM},
\begin{align}
Z^\text{pert}(N)=e^AC^{-\frac{1}{3}}\Ai\Bigl[C^{-\frac{1}{3}}(N-B)\Bigr],
\label{Airy}
\end{align}
with some constants $A$ \cite{KEK}, $B$ and $C$, up to non-perturbative corrections in $1/N$.
Later it was pointed out that the partition function of this theory can be rewritten as that of an ideal Fermi gas system with $N$ particles \cite{MP}.
This formalism is so efficient that it not only provided a simple rederivation of the above result, but even allowed the exact analysis of the non-perturbative corrections \cite{HMO1,PY,HMO2,CM,HMO3,HMMO}.

In \cite{MP} the authors also showed that the Fermi gas formalism works for theories of general ${\widehat A}$-type quivers, and that the completion by an Airy function is universal for these theories.
The superconformal Chern-Simons theories of the ${\widehat A}$-type quivers were also studied in detail, including the perturbative coefficients $A,B,C$ and various non-perturbative corrections \cite{HM,MN1,MN2,MN3}.\footnote{
The $N_f$ flavor matrix model also allows the Fermi gas formalism which turns out to be equivalent to that for some of the ${\widehat A}$-type quiver.
These matrix models were studied in \cite{MePu,GM,HaOk}.
The equivalence between these matrix models is proven systematically in \cite{DF}.
}

It was further conjectured in \cite{MP} that the completion by an Airy function is universal even for other theories of the M2-branes.
However, in the case of the ${\widehat D}{\widehat E}$-type quivers, it was not trivial whether the universal behaviour of the Airy function is valid because of the lack of the Fermi gas formalism due to its non-circular structure and the non-uniform comarks.

\begin{figure}[ht!]
\centering
\includegraphics[width=10cm]{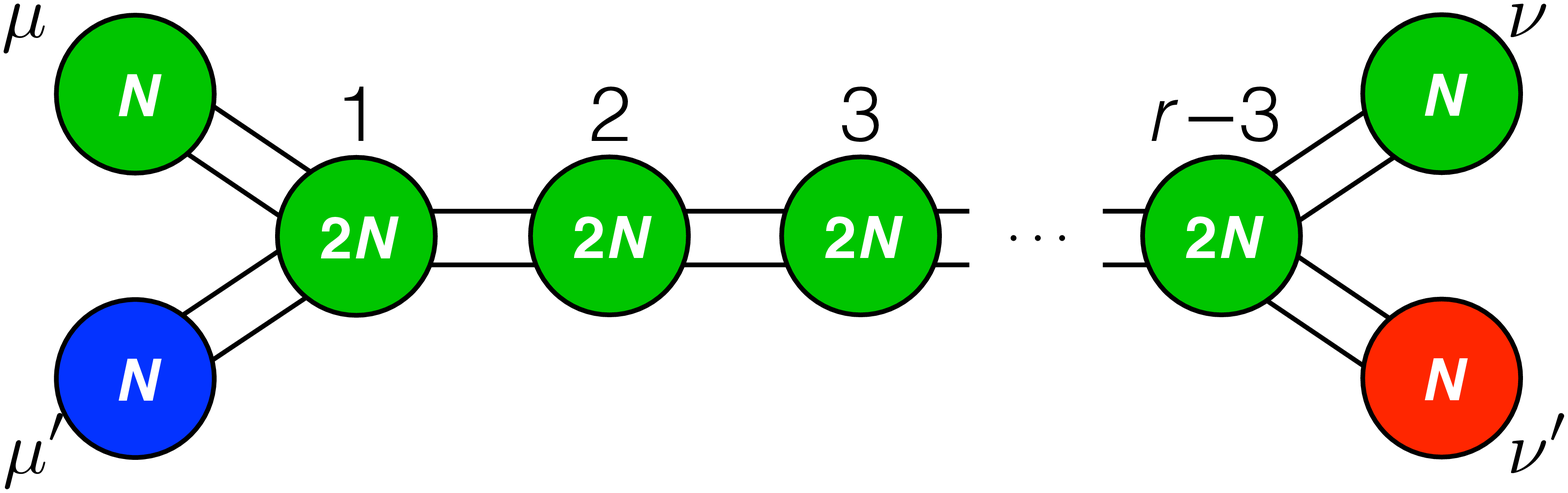}
\caption{The label of the vertices of the ${\widehat D}_r$ quiver.
The number on each vertex is the rank of the gauge group, which is proportional to the comark.}
\label{Dquiver}
\end{figure}

In this paper we consider general ${\widehat D}$-type quivers, and obtain a positive answer to this question.
Here we shall explain our setup.
We consider the quiver superconformal Chern-Simons theory whose quiver is the ${\widehat D}_r$ Dynkin diagram.
We call the two vertices on the left end as $\mu$ and $\mu^\prime$, those on the right end as $\nu$ and $\nu^\prime$, and label the others as $1,2,\cdots, r-3$, as in figure \ref{Dquiver}.
In addition to the number $N_m$ of the gauge group $U(N_m)$, the Chern-Simons level is assigned on each vertex, which we take as $(m=1,2,\cdots,r-3)$
\begin{align}
k_\mu&=k(s_1-s_2),&
k_m&=k(s_{m+1}-s_{m+2}),&
k_\nu&=k(s_{r-1}-s_{r}),\nonumber\\
k'_\mu&=k(-s_1-s_2),&&&
k'_\nu&=k(s_{r-1}+s_{r}),
\label{levelsforD}
\end{align}
with $s_m$ extended to arbitrary real numbers.
This choice is the most general one under the requirement of the superconformal symmetry of the theory \cite{GAH}.
Below, we shall first present a Fermi gas description for the partition function of this theory in section \ref{Fermigas}.
In the Fermi gas formalism, it is rather convenient to introduce the chemical potential $\mu$ dual to $N$ and study the grand potential $J(\mu)$ defined by
\begin{align}
e^{J(\mu)}=\sum_{N=0}^\infty e^{\mu N}Z(N).
\label{JZ}
\end{align}
Other than $s_m$, the grand potential is controlled by two parameters, the chemical potential $\mu$ and the overall Chern-Simons level $k$ which plays the role of the Planck constant of this quantum statistical system, $\hbar=2\pi k$.
The grand potential turns out to be a cubic polynomial of $\mu$ if we neglect the non-perturbative effects.
In this manner, all order perturbative corrections to the partition function in $1/N$ are taken into account.
As a result, we obtain the expression of an Airy function \eqref{Airy} in section \ref{BC}.
The coefficient $C$ relevant to the leading $N^{\frac{3}{2}}$ behaviour is obtained as
\begin{align}
C=\frac{1}{\pi^2k}
\biggl(\frac{1}{\sigma_0\sigma_1}
+\sum_{m=1}^{r-1}\frac{s_m-s_{m+1}}{\sigma_m\sigma_{m+1}}
+\frac{s_r}{\sigma_r\sigma_{r+1}}\biggr),
\label{C}
\end{align}
where the variables $\sigma$ are given by
\begin{align}
\sigma_m=\sum_{n=1}^r(|s_m-s_n|+|s_m+s_n|)-4|s_m|,\quad
\sigma_0=2(r-2),\quad\sigma_{r+1}=2\sum_{n=1}^r|s_n|,
\label{sigma&s}
\end{align}
with the reordered $s_m$,
\begin{align}
0\le |s_r|\le |s_{r-1}|\le \cdots \le |s_1|.
\label{sorder}
\end{align}
This coincides perfectly with the previous proposal in \cite{CHJ}, where the authors further tried to give a Fermi surface interpretation to their proposal.  
In the limit of $k\rightarrow 0$, we can also compute the coefficient $B$, and the result is
\begin{align}
B=\frac{\pi^2C}{3}-\frac{1}{6k}
\biggl(\sum_{m=1}^r\frac{1}{\sigma_m}+\frac{1}{\sigma_{r+1}}\biggr)
+{\cal O}(k).
\label{B}
\end{align}

If we further restrict ourselves to the special values of $s_m$
\begin{align}
s_1=s_2=\cdots=s_r=1,
\label{Dspecial}
\end{align}
we can compute the higher order corrections in $k$ to the grand potential, including the perturbative coefficients and the non-perturbative corrections.
We analyse this model in section \ref{specialinst} and conclude that, for the non-perturbative corrections, there are three kinds of membrane instantons with exponents $e^{-\frac{2\mu}{r}}$, $e^{-\frac{2\mu}{r-2}}$ and $e^{-\frac{\mu}{2}}$.
For the coefficient $A$, we observe that the result is consistent with
\begin{align}
A=\frac{1}{2}
\Bigl[A_{\text{ABJM}}(2rk)+r^2A_{\text{ABJM}}(2(r-2)k)\Bigr],
\label{Aspecial}
\end{align}
up to ${\cal O}(k^5)$, where $A_{\text{ABJM}}(k)$ denotes that coefficient for the ABJM theory.
Both the coefficient $A$ and the membrane instantons are reminiscent of the expressions for the ${\cal N}=4$ supersymmetric theories of the ${\widehat A}$-type quiver \cite{MN1,MN2}.

\section{Fermi gas formalism}\label{Fermigas}
In this section, we shall present a Fermi gas description for the superconformal Chern-Simons theories of the $\widehat D$-type quiver.

The partition function of this theory is given by
\begin{align}
Z(N)=
\int\frac{D^N\mu}{N!}\frac{D^N\mu'}{N!}
\frac{D^N\nu}{N!}\frac{D^N\nu'}{N!}
\prod_{m=1}^{r-3}\frac{D^{2N}\lambda^{(m)}}{(2N)!}\frac{V}{H},
\label{Z}
\end{align}
with the integration measure
\begin{align}
D\mu_i&=\frac{d\mu_i}{2\pi}e^{\frac{ik_\mu}{4\pi}(\mu_i)^2},&
D\lambda^{(m)}_a&=\frac{d\lambda^{(m)}_a}{2\pi}
e^{\frac{ik_{m}}{4\pi}(\lambda^{(m)}_a)^2},&
D\nu_i&=\frac{d\nu_i}{2\pi}e^{\frac{ik_\nu}{4\pi}(\nu_i)^2},
\nonumber\\
D\mu'_i&=\frac{d\mu'_i}{2\pi}e^{\frac{ik'_\mu}{4\pi}(\mu_i')^2},&&&
D\nu'_i&=\frac{d\nu'_i}{2\pi}e^{\frac{ik'_\nu}{4\pi}(\nu_i')^2}.
\label{int}
\end{align}
Here the numerator $V$, coming from the vector multiplets in the adjoint representation, is given by
\begin{align}
V&=\prod_{i<j}^N\Bigl(2\sinh\frac{\mu_i-\mu_j}{2}\Bigr)^2
\prod_{i<j}^N\Bigl(2\sinh\frac{\mu'_i-\mu'_j}{2}\Bigr)^2
\prod_{m=1}^{r-3}\prod_{a<b}^{2N}
\Bigl(2\sinh\frac{\lambda^{(m)}_a-\lambda^{(m)}_b}{2}\Bigr)^2
\nonumber\\&\qquad\times
\prod_{i<j}^N\Bigl(2\sinh\frac{\nu_i-\nu_j}{2}\Bigr)^2
\prod_{i<j}^N\Bigl(2\sinh\frac{\nu'_i-\nu'_j}{2}\Bigr)^2,
\label{V}
\end{align}
and the denominator $H$, coming from the hypermultiplets in the bifundamental representation, is
\begin{align}
H&=\prod_{i,a}^{N,2N}\Bigl(2\cosh\frac{\mu_i-\lambda^{(1)}_a}{2}\Bigr)
\prod_{i,a}^{N,2N}\Bigl(2\cosh\frac{\mu'_i-\lambda^{(1)}_a}{2}\Bigr)
\prod_{m=1}^{r-4}\prod_{a,b}^{2N,2N}
\Bigl(2\cosh\frac{\lambda^{(m)}_a-\lambda^{(m+1)}_b}{2}\Bigr)
\nonumber\\&\qquad\times
\prod_{a,i}^{2N,N}\Bigl(2\cosh\frac{\lambda^{(r-3)}_a-\nu_i}{2}\Bigr)
\prod_{a,i}^{2N,N}\Bigl(2\cosh\frac{\lambda^{(r-3)}_a-\nu'_i}{2}\Bigr).
\label{H}
\end{align}

\subsection{Density matrix from matrix model}
To express the partition function \eqref{Z} of the superconformal Chern-Simons matrix model of the $\widehat D$-type quiver in terms of that of a Fermi gas system, in this subsection let us first rewrite the generating function of the matrix model into a Fredholm determinant.

First, we rewrite the integrand of the matrix model \eqref{Z} into\footnote{We knew of a related work \cite{ADF} from the reference list of \cite{DF}.
In a seminar by Nadav Drukker at Nagoya university, we learned that actually they had a similar idea in rewriting the integration measure into a determinant of hyperbolic cosecant functions as in \eqref{V/H}.
We are grateful to Nadav Drukker for valuable discussions.}\footnote{
Since the singularities appearing in the first and last determinants are originally absent in \eqref{H}, we expect them to be harmless.}
\begin{align}
\frac{V}{H}
&=\det\biggl(\frac{1}{2\sinh\frac{\mu_i-\mu'_j}{2}}\biggr)_{N\times N}
\det
\biggl(\frac{1}{2\cosh\frac{\overline\mu_a-\lambda^{(1)}_b}{2}}\biggr)
_{2N\times 2N}
\prod_{m=1}^{r-4}\det
\biggl(\frac{1}{2\cosh\frac{\lambda^{(m)}_a-\lambda^{(m+1)}_b}{2}}\biggr)
_{2N\times 2N}
\nonumber\\&\quad\times
\det
\biggl(\frac{1}{2\cosh\frac{\lambda^{(r-3)}_a-\overline\nu_b}{2}}\biggr)
_{2N\times 2N}
\det\biggl(\frac{1}{2\sinh\frac{\nu_i-\nu'_j}{2}}\biggr)_{N\times N}.
\label{V/H}
\end{align}
Here we have introduced the combined variables $(\overline\mu)_{a=1,\cdots,2N}=\bigl((\mu)_{a=1,\cdots,N},(\mu')_{a-N=1,\cdots,N}\bigr)$ and $(\overline\nu)_{a=1,\cdots,2N}=\bigl((\nu)_{a=1,\cdots,N},(\nu')_{a-N=1,\cdots,N}\bigr)$.
Namely, the second factor in \eqref{V/H} is the determinant of a matrix which is a vertical array of two $N\times 2N$ rectangular matrices, one with components $\bigl(2\cosh\frac{\mu_i-\lambda^{(1)}_b}{2}\bigr)^{-1}$ and the other with components $\bigl(2\cosh\frac{\mu'_i-\lambda^{(1)}_b}{2}\bigr)^{-1}$.
Similarly, the second last factor is the determinant of a matrix which is a horizontal array of two $2N\times N$ rectangular matrices, $\bigl(\bigl(2\cosh\frac{\lambda^{(r-3)}_a-\nu_i}{2}\bigr)^{-1}\bigr)_{2N\times N}$ and $\bigl(\bigl(2\cosh\frac{\lambda^{(r-3)}_a-\nu'_i}{2}\bigr)^{-1}\bigr)_{2N\times N}$.
Note that in the above rewriting we have used the formulae
\begin{align}
\frac{\prod_{i<j}2\sinh\frac{\mu_i-\mu_j}{2}
\prod_{i<j}2\sinh\frac{\nu_i-\nu_j}{2}}
{\prod_{i,j}2\cosh\frac{\mu_i-\nu_j}{2}}
&=\det\frac{1}{2\cosh\frac{\mu_i-\nu_j}{2}},\nonumber\\
\frac{\prod_{i<j}^N2\sinh\frac{\mu_i-\mu_j}{2}
\prod_{i<j}^N2\sinh\frac{\nu_i-\nu_j}{2}}
{\prod_{i,j}2\sinh\frac{\mu_i-\nu_j}{2}}
&=(-1)^{\frac{1}{2}(N-1)N}\det\frac{1}{2\sinh\frac{\mu_i-\nu_j}{2}},
\end{align}
which follow from the Cauchy identity
\begin{align}
\frac{\prod_{i<j}(x_i-x_j)\prod_{i<j}(y_i-y_j)}{\prod_{i,j}(x_i+y_j)}
=\det\frac{1}{x_i+y_j},
\label{Cauchy}
\end{align}
by the substitutions $x_i=e^{\mu_i}$ and $y_j=e^{\nu_j}$ or $y_j=-e^{\nu_j}$.

Then, using the formula proved in appendix A of \cite{MM} with the same ranks, we can combine the series of $2N\times 2N$ determinants into
\begin{align}
Z(N)=\int\frac{D^N\mu}{N!}\frac{D^N\mu'}{N!}
\frac{D^N\nu}{N!}\frac{D^N\nu'}{N!}
\det M(\mu_i,\mu'_j)
\det\overline L(\overline\mu_a,\overline\nu_b)
\det N(\nu_i,\nu'_j).
\label{MLN}
\end{align}
Here the functions $M$, $N$ and $\overline L$ denote\footnote{The rank $N$ of the gauge group should not be confused with the function $N(\nu,\nu')$.
Also, the chemical potential $\mu$ appearing later should not be confused with the integration variables $\mu_i$ and $\mu'_i$.
}
\begin{align}
&M(\mu_i,\mu'_j)=\frac{1}{2\sinh\frac{\mu_i-\mu'_j}{2}},\quad
N(\nu_i,\nu'_j)=\frac{1}{2\sinh\frac{\nu_i-\nu'_j}{2}},\nonumber\\
&\overline L(\overline\mu_a,\overline\nu_b)
=\int\prod_{m=1}^{r-3}D\lambda^{(m)}
\frac{1}{2\cosh\frac{\overline\mu_a-\lambda^{(1)}}{2}}
\biggl[\prod_{m=1}^{r-4}
\frac{1}{2\cosh\frac{\lambda^{(m)}-\lambda^{(m+1)}}{2}}\biggr]
\frac{1}{2\cosh\frac{\lambda^{(r-3)}-\overline\nu_b}{2}},
\end{align}
with the matrix in the second determinant in \eqref{MLN} given explicitly by
\begin{align}
\overline L(\overline\mu_a,\overline\nu_b)
=\begin{pmatrix}
L(\mu_i,\nu_j)&L(\mu_i,\nu'_j)\\
L(\mu'_i,\nu_j)&L(\mu'_i,\nu'_j)
\end{pmatrix}.
\end{align}
Furthermore, if we use the formula in \cite{MM} for different ranks, we can perform the $\mu'$ and $\nu'$ integrations to find
\begin{align}
Z(N)=(-1)^N\int\frac{D^N\mu}{N!}\frac{D^N\nu}{N!}
\det\begin{pmatrix}
L(\mu_i,\nu_j)&(L\bullet N)(\mu_i,\nu_j)\\
(M\bullet L)(\mu_i,\nu_j)&(M\bullet L\bullet N)(\mu_i,\nu_j)
\end{pmatrix},
\end{align}
where $\bullet$ stands for either the $\mu'$ integration or the $\nu'$ integration in \eqref{int}.
It should be clear from the context which integration it stands for.
For example, $(M\bullet L\bullet N)(\mu_i,\nu_j)$ in the lower-right block denotes
\begin{align}
(M\bullet L\bullet N)(\mu_i,\nu_j)
=\int D\mu'D\nu'M(\mu_i,\mu')L(\mu',\nu')N(\nu',\nu_j).
\end{align}

Now we can apply the formula \eqref{pfaffian} in our appendix \ref{detpf} and find
\begin{align}
Z(N)=(-1)^{N+\frac{1}{2}(N-1)N}\int\frac{D^N\mu}{N!}
\pf\overline P,\quad\overline P=\begin{pmatrix}P_{11}&P_{12}\\P_{21}&P_{22}\end{pmatrix},
\end{align}
with the four $N\times N$ blocks given by
\begin{align}
P_{11}
&=-(L\circ N\bullet L+L\bullet N\circ L)(\mu_i,\mu_j),
&P_{12}
&=(L\circ N\bullet L+L\bullet N\circ L)\bullet M(\mu_i,\mu_j),
\nonumber\\
P_{21}
&=-M\bullet
(L\circ N\bullet L+L\bullet N\circ L)(\mu_i,\mu_j),
&P_{22}
&=M\bullet
(L\circ N\bullet L+L\bullet N\circ L)\bullet M(\mu_i,\mu_j).
\end{align}
Here $\circ$ denotes the $\nu$ integration.
If we further introduce the chemical potential $\mu$ and define the grand potential $J(\mu)$ as \eqref{JZ}, we find
\begin{align}
e^{J(\mu)}=\sqrt{\det(\overline I+e^\mu\overline\rho)},
\label{det}
\end{align}
by using the formula in appendix \ref{squareroot}.
Here the density matrix $\overline\rho$ is
\begin{align}
\overline\rho=\overline\Omega\,\overline P,
\end{align}
and the other matrices are given in \eqref{OmegaI}.
Then it is not difficult to observe that the density matrix can be put into
\begin{align}
\overline\rho=\begin{pmatrix}-M_{\mu\mu'}&0\\0&I\end{pmatrix}
\begin{pmatrix}L_{\mu'\nu}&L_{\mu'\nu'}\\
L_{\mu\nu}&L_{\mu\nu'}\end{pmatrix}
\begin{pmatrix}N_{\nu\nu'}&0\\0&N_{\nu'\nu}\end{pmatrix}
\begin{pmatrix}L_{\nu'\mu}&L_{\nu'\mu'}\\
L_{\nu\mu}&L_{\nu\mu'}\end{pmatrix}
\begin{pmatrix}I&0\\0&-M_{\mu'\mu}\end{pmatrix}.
\end{align}
Here we have regarded the functions $M$, $N$ and $L$ as matrices, and contracted the adjacent indices by integrations \eqref{int}, without displaying $\bullet$ or $\circ$ explicitly.
After suitable similarity transformations and rearrangements, we can further put the density matrix into
\begin{align}
\overline\rho=-\begin{pmatrix}L_{\nu\mu}&L_{\nu\mu'}\\
L_{\nu'\mu}&L_{\nu'\mu'}\end{pmatrix}
\begin{pmatrix}0&M_{\mu\mu'}\\M_{\mu'\mu}&0\end{pmatrix}
\begin{pmatrix}L_{\mu\nu}&L_{\mu\nu'}\\
L_{\mu'\nu}&L_{\mu'\nu'}\end{pmatrix}
\begin{pmatrix}0&N_{\nu\nu'}\\N_{\nu'\nu}&0\end{pmatrix}.
\label{rho22}
\end{align}

\subsection{Operator formalism for density matrix}

In the previous subsection, we have reduced the study of the superconformal Chern-Simons matrix model of the $\widehat D$-type quiver into an integration kernel $\overline\rho$.
To express the superconformal Chern-Simons matrix model in terms of a Fermi gas system with $N$ particles, we need to further rewrite the integration kernel in the operator formalism.
For this purpose, we introduce the canonical coordinate/momentum operators ${\widehat q}$ and ${\widehat p}$ obeying the canonical commutation relation
\begin{align}
[{\widehat q},{\widehat p}]=i\hbar,
\label{ccr}
\end{align}
with the Planck constant given by $\hbar=2\pi k$ and normalize the coordinate eigenstate as
\begin{align}
\langle q|q'\rangle=2\pi\delta(q-q').
\end{align}

It is not difficult to spell out each block in $\overline\rho$ \eqref{rho22} explicitly in terms of the hyperbolic functions and the integrations.
After rescaling the integration variables by $1/k$, we find that the density matrix is given by
\begin{align}
\overline\rho=\begin{pmatrix}
\langle\nu|\widehat\rho_+|\nu\rangle&
\langle\nu|\widehat\rho_-|\nu'\rangle\\
\langle\nu'|\widehat\rho_+|\nu\rangle&
\langle\nu'|\widehat\rho_-|\nu'\rangle
\end{pmatrix},
\label{rhobar}
\end{align}
with the operators $\widehat\rho_\pm$
\begin{align}
\widehat\rho_+&=
\frac{1}{2\cosh\frac{\widehat p}{2}}
e^{-\frac{i}{2\hbar}(s_{r-1}-s_{r-2})\widehat q^2}
\frac{1}{2\cosh\frac{\widehat p}{2}}
e^{-\frac{i}{2\hbar}(s_{r-2}-s_{r-3})\widehat q^2}\cdots
e^{-\frac{i}{2\hbar}(s_3-s_2)\widehat q^2}
\frac{1}{2\cosh\frac{\widehat p}{2}}
\nonumber\\&\quad\times
\biggl(e^{-\frac{i}{2\hbar}(s_2-s_1)\widehat q^2}
\frac{\tanh\frac{\widehat p}{2}}{2}
e^{-\frac{i}{2\hbar}(s_1+s_2)\widehat q^2}
+e^{-\frac{i}{2\hbar}(s_2+s_1)\widehat q^2}
\frac{\tanh\frac{\widehat p}{2}}{2}
e^{\frac{i}{2\hbar}(s_1-s_2)\widehat q^2}\biggr)
\nonumber\\&\quad\times
\frac{1}{2\cosh\frac{\widehat p}{2}}
e^{\frac{i}{2\hbar}(s_2-s_3)\widehat q^2}\cdots
e^{\frac{i}{2\hbar}(s_{r-2}-s_{r-1})\widehat q^2}
\frac{1}{2\cosh\frac{\widehat p}{2}}
e^{\frac{i}{2\hbar}(s_{r-1}+s_{r})\widehat q^2}
\frac{\tanh\frac{\widehat p}{2}}{2}
e^{-\frac{i}{2\hbar}(s_{r}-s_{r-1})\widehat q^2},\nonumber\\
\widehat\rho_-&=
\frac{1}{2\cosh\frac{\widehat p}{2}}
e^{-\frac{i}{2\hbar}(s_{r-1}-s_{r-2})\widehat q^2}
\frac{1}{2\cosh\frac{\widehat p}{2}}
e^{-\frac{i}{2\hbar}(s_{r-2}-s_{r-3})\widehat q^2}\cdots
e^{-\frac{i}{2\hbar}(s_3-s_2)\widehat q^2}
\frac{1}{2\cosh\frac{\widehat p}{2}}
\nonumber\\&\quad\times
\biggl(e^{-\frac{i}{2\hbar}(s_2-s_1)\widehat q^2}
\frac{\tanh\frac{\widehat p}{2}}{2}
e^{-\frac{i}{2\hbar}(s_1+s_2)\widehat q^2}
+e^{-\frac{i}{2\hbar}(s_2+s_1)\widehat q^2}
\frac{\tanh\frac{\widehat p}{2}}{2}
e^{\frac{i}{2\hbar}(s_1-s_2)\widehat q^2}\biggr)
\nonumber\\&\quad\times
\frac{1}{2\cosh\frac{\widehat p}{2}}
e^{\frac{i}{2\hbar}(s_2-s_3)\widehat q^2}\cdots
e^{\frac{i}{2\hbar}(s_{r-2}-s_{r-1})\widehat q^2}
\frac{1}{2\cosh\frac{\widehat p}{2}}
e^{\frac{i}{2\hbar}(s_{r-1}-s_{r})\widehat q^2}
\frac{\tanh\frac{\widehat p}{2}}{2}
e^{\frac{i}{2\hbar}(s_{r}+s_{r-1})\widehat q^2}.
\label{rho+-}
\end{align}
In the derivation we have used the formulae
\begin{align}
\langle q|\frac{1}{2\cosh\frac{\widehat p}{2}}|q'\rangle
=\frac{1}{k}\frac{1}{2\cosh\frac{q-q'}{2k}},\quad
\langle q|\frac{\tanh\frac{\widehat p}{2}}{2}|q'\rangle
=\frac{i}{k}\frac{1}{2\sinh\frac{q-q'}{2k}}.
\end{align}

Using this result we can simplify the grand potential \eqref{det}.
In \eqref{det} the determinant is taken simultaneously over the functional space (or, in the operator formalism, the phase space) and over the two dimensional space.
However, since the left two components and the right two components in \eqref{rhobar} are identical, the determinant over the two dimensional space can be taken trivially
\begin{align}
e^{J(\mu)}=\sqrt{\det(I+e^\mu\widehat\rho)},
\label{Jrhorelation}
\end{align}
with the density matrix purely in the phase space given by $\widehat\rho=\widehat\rho_++\widehat\rho_-$.
After performing the similarity transformation to move $e^{\frac{i}{2\hbar}s_{r-1}\widehat q^2}$ on the right end of $\widehat\rho_\pm$ in \eqref{rho+-} into the left end, we can use the relation
\begin{align}
e^{\frac{i}{2\hbar}s\widehat q^2}F(\widehat p)
e^{-\frac{i}{2\hbar}s\widehat q^2}
=F(\widehat p-s\widehat q),
\end{align}
to rewrite the density matrix into
\begin{align}
\widehat\rho
&=\frac{1}{2\cosh\frac{{\widehat p}-s_{r-1}{\widehat q}}{2}}\cdots
\frac{1}{2\cosh\frac{{\widehat p}-s_2{\widehat q}}{2}}
\frac{\tanh\frac{{\widehat p}-s_1{\widehat q}}{2}
+\tanh\frac{{\widehat p}+s_1{\widehat q}}{2}}{2}\nonumber \\
&\quad\times
\frac{1}{2\cosh\frac{{\widehat p}+s_2{\widehat q}}{2}}\cdots
\frac{1}{2\cosh\frac{{\widehat p}+s_{r-1}{\widehat q}}{2}}
\frac{\tanh\frac{{\widehat p}-s_r{\widehat q}}{2}
+\tanh\frac{{\widehat p}+s_r{\widehat q}}{2}}{2}.
\label{rho}
\end{align}

\section{Large $N$ behaviour from Fermi surface analysis} \label{BC}
In the previous section, after switching from the partition function $Z(N)$ to the grand potential $J(\mu)$, we find that $J(\mu)$ is expressed in terms of the Fredholm determinant \eqref{Jrhorelation} of the density matrix \eqref{rho}.
Since the relation is very similar to the case of the $\widehat A$-type quiver \cite{MP}, we expect that the perturbative corrections to the partition function again sum up to an Airy function as \eqref{Airy} with some constants $C$, $B$ and $A$.
Indeed, the expression of the Airy function follows from the large $E$ behaviour of the number of states with energy smaller than $E$ \cite{MP}\footnote{
The overall factor $2$ compared with the case of the ${\widehat A}$-type quivers is due to the square-root in \eqref{Jrhorelation}.
}
\begin{align}
n(E)=\tr[\theta(E-\log{\widehat \rho}^{-1})]
=2\biggl(CE^2+B-\frac{\pi^2C}{3}\biggr)+{\cal O}(e^{-E}).\label{npert}
\end{align}
In this section we shall show this relation, with explicit expressions of $C$ and $B$ up to ${\cal O}(k)$, by the technique used in \cite{MP}.

\begin{figure}[ht!]
\centering
\includegraphics[width=7cm]{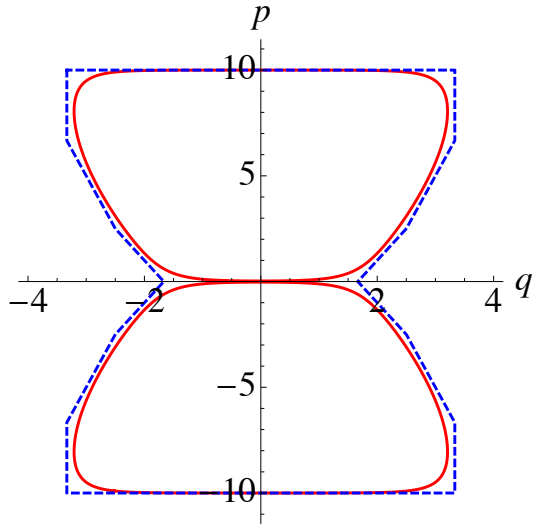}
\caption{The classical Fermi surface for $\{s_m\}=\{1,2,3\}$ at $E=10$ (the solid red line) and the polygon to which the Fermi surface approaches in the limit of $E\rightarrow\infty$ (the dashed blue line).}
\label{fermisurface}
\end{figure}

For this purpose, let us consider the classical limit of the Fermi gas system ($\hbar\rightarrow 0$).
Classically, the number of states $n(E)$ is given by the phase space volume
\begin{align}
n(E)=\frac{1}{2\pi\hbar}
\vol\{(q,p)\in\mathbb{R}^2|\log\rho_0^{-1}\le E\},
\end{align}
where $\rho_0$ is the classical density matrix obtained from $\widehat\rho$ \eqref{rho} by neglecting the commutators
\begin{align}
\log(\rho_0(q,p))^{-1}
=\sum_{m=1}^r\biggl[\log\Bigl(2\cosh\frac{p-s_mq}{2}\Bigr)
+\log\Bigl(2\cosh\frac{p+s_mq}{2}\Bigr)\biggr]-\log(2\sinh p)^2.
\label{logrho0}
\end{align}
Since the right-hand side is independent of the signs and the ordering of $s_m$, in this section we assume $s_m\ge 0$ and \eqref{sorder} without loss of generality.
The classical Fermi surface (i.e.\ the boundary of the region with $\log\rho_0^{-1}\le E$) is plotted in figure \ref{fermisurface}.
Since the region is symmetric under the reflections $q\mapsto -q$ and $p\mapsto -p$, below we consider only the subregion in the first quadrant ${\mathbb R}_{\ge 0}^2$.
We can further divide the volume $\vol\{(q,p)\in{\mathbb R}_{\ge 0}^2|\log\rho_0^{-1}\le E\}$ into the leading contribution in the limit of large $E$, $V_\text{pol}$, and the deviation from it, $\delta V$, as
\begin{align}
n(E)=\frac{2}{\pi\hbar}(V_{\text{pol}}-\delta V).
\label{nE}
\end{align}
The main contribution $V_{\text{pol}}$ can be computed by approximating the hyperbolic functions by rational functions 
\begin{align}
V_\text{pol}=\vol\biggl\{(q,p)\in{\mathbb R}_{\ge 0}^2\bigg|\sum_{m=1}^r\biggl[\frac{|p-s_mq|}{2}+\frac{|p+s_mq|}{2}\biggr]-2|p|\le E\biggr\}.
\end{align}
Since the above subregion on the first quadrant is a polygon (see figure \ref{fermisurface}), whose vertices are located at
\begin{align}
\biggl(0,\frac{2E}{\sigma_0}\biggr),\quad
\biggl(\frac{2E}{\sigma_m},\frac{2Es_m}{\sigma_m}\biggr),\quad
\biggl(\frac{2E}{\sigma_{r+1}},0\biggr),
\end{align}
with $\sigma$ given by $s_m$ as in \eqref{sigma&s}, the volume of this subregion is
\begin{align}
V_{\text{pol}}
&=2E^2
\biggl(\frac{1}{\sigma_0\sigma_1}
+\sum_{m=1}^{r-1}\frac{s_m-s_{m+1}}{\sigma_m\sigma_{m+1}}
+\frac{s_r}{\sigma_r\sigma_{r+1}}\biggr).
\label{Vpol}
\end{align}

\begin{figure}[ht!]
\centering
\includegraphics[width=7cm]{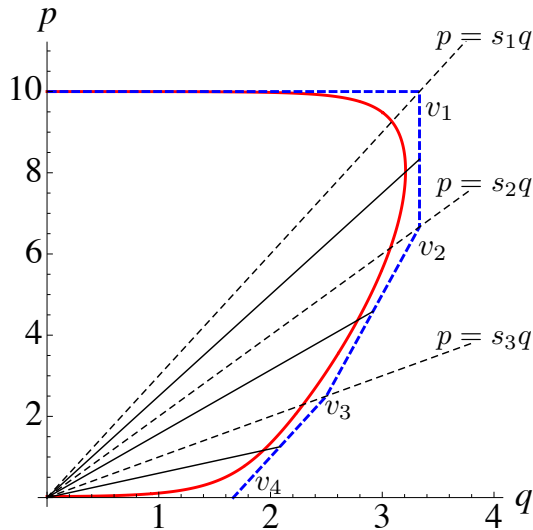}
\caption{
The region contributing to $v_m$ is the region surrounded by the classical Fermi surface (the solid red line), the boundary of the polygon (the dashed blue line) and the solid black lines next to the line of $p=s_mq$.
Each solid black line connects the origin and the midpoint on the edge of the polygon.
}
\label{fsdeviation}
\end{figure}

Now we consider the deviation from the volume of the limit polygon, $\delta V$.
First we divide the region between the classical Fermi surface and the  polygon into the pieces around each line of $p=s_mq$ and $p=0$, as in figure \ref{fsdeviation}, and call the volume of each piece $v_m$ and $v_{r+1}$ respectively,
\begin{align}
\delta V=\sum_{m=1}^rv_m+v_{r+1}.
\label{deltaV}
\end{align}
Assuming that $q$ and $p$ are of order $E$ on the classical Fermi surface, we can approximate its segment near the line $p=s_mq$ as
\begin{align}
\sum_{\ell\neq m}
\biggl[\frac{|p-s_\ell q|}{2}+\frac{|p+s_\ell q|}{2}\biggr]
+\log 2\cosh\frac{p-s_mq}{2}+\frac{|p+s_mq|}{2}-2|p|=E,
\label{approx}
\end{align}
if we neglect the non-perturbative ${\cal O}(e^{-E})$ corrections.
To calculate $v_m$, it is convenient to introduce the tilted coordinate $({\widetilde q},{\widetilde p})=(q,p-s_mq)$,
\begin{align}
v_m=\int_{{\widetilde p}_-}^{{\widetilde p}_+}d{\widetilde p}(q({\widetilde p})-q^\prime({\widetilde p})),
\end{align}
where $q({\widetilde p})$ and $q^\prime({\widetilde p})$ are the ${\widetilde q}$-coordinate of a point $({\widetilde q},{\widetilde p})$ on the limiting polygon of the Fermi surface and that on the approximant \eqref{approx}, and ${\widetilde p}_\pm$ are the midpoints on the edges of the polygon.
Noting $p-s_\ell q<0$ for $\ell<m$ and $p-s_\ell q>0$ for $\ell>m$ in this piece, we can compute $v_m$ as
\begin{align}
v_m=\int_{{\widetilde p}_-}^{{\widetilde p}_+}d{\widetilde p}
\frac{2}{\sigma_m}
\Bigl(\log2\cosh\frac{\widetilde p}{2}-\frac{|{\widetilde p}|}{2}\Bigr)
\simeq\frac{\pi^2}{3\sigma_m}.
\label{vm}
\end{align}
Here, although originally the integral interval, $[{\widetilde p}_-,{\widetilde p}_+]$ is finite, we can replace it with the whole real axis $(-\infty,\infty)$ without affecting the perturbative behaviour in \eqref{npert}, since the integrand is exponentially small for large ${\widetilde p}$.
The contribution from the piece around $p=0$ can be calculated similarly, 
\begin{align}
v_{r+1}=\frac{\pi^2}{3\sigma_{r+1}}.
\label{vr+1}
\end{align}
Substituting these results \eqref{Vpol}, \eqref{deltaV}, \eqref{vm}, \eqref{vr+1} into \eqref{nE}, we obtain the large $E$ expression of $n(E)$ \eqref{npert} with $C$ and $B$ given by \eqref{C} and \eqref{B}.

So far we have been neglecting the quantum corrections.
Though it is difficult to take care of them due to the variety of arguments of the hyperbolic functions in the density matrix \eqref{rho}, we can make the following estimation.
There are two kinds of $\hbar$-corrections, the Wigner transformation of each operator and the commutators of operators coming from the Baker-Campbell-Hausdorff formula.
According to the Wigner transformation formula, the former corrections always start with the second derivatives of each term in \eqref{logrho0}.
Also, if the Hamiltonian is hermitian, since there are only nested commutators, the latter corrections again start with the second derivatives.\footnote{The requirement of hermiticity is essential also in the discussion in the $\widehat A$-type quiver \cite{MP,MN1}.}
Therefore, since the second derivatives of the hyperbolic functions are always exponentially suppressed, the quantum corrections never change the asymptotic polygon of the Fermi surface in the limit of $E\rightarrow\infty$.
This ensures the behaviour of $n(E)$ \eqref{npert} with $C$ uncorrected, and therefore that the perturbative partition function is given as an Airy function even with all order quantum corrections.
On the other hand, $B$ is possibly corrected due to the quantum effect.

\section{$A$ and instantons for special quivers}\label{specialinst}
In this section, we restrict ourselves to the cases where the Chern-Simons levels are given by \eqref{levelsforD} with a uniform value of $s_m$ \eqref{Dspecial}, where we set that value to be $1$, which is always possible by the redefinition of $k$.
Under this restriction the exact large $\mu$ expansion of the grand potential can be computed systematically order by order in $k$.
As a result, we obtain the constant part $A$ (which appear in the partition function as \eqref{Airy}) and the non-perturbative corrections (${\cal O}(e^{-\mu})$) in the grand potential.
These are just what we did in the theory of the ${\widehat A}$-type quiver with ${\cal N}=4$ supersymmetry enhancement \cite{MP,MN1,MN2}.

For the simplicity of explanation, let us define the Hamiltonian, $e^{-{\widehat H}}={\widehat\rho}$.
After a similarity transformation, the Hamiltonian is given explicitly by
\begin{align}
e^{-{\widehat H}}=
e^{-\frac{r-2}{2}{\widehat U}}
e^{\widehat S}
e^{-(r-2){\widehat T}}
e^{\widehat S}
e^{-\frac{r-2}{2}{\widehat U}},
\label{eHeUeSeT}
\end{align}
where we have introduced new variables
\begin{align}
{\widehat U}=\log 2\cosh\frac{\widehat Q}{2},\quad
{\widehat T}=\log 2\cosh\frac{\widehat P}{2},\quad
{\widehat S}=\log \frac{\tanh\frac{\widehat Q}{2}+\tanh\frac{\widehat P}{2}}{2},
\label{UTS}
\end{align}
with ${\widehat Q}={\widehat p}+{\widehat q}$ and ${\widehat P}={\widehat p}-{\widehat q}$.
Note that the Planck constant is doubled in the new canonical variables, $[{\widehat Q},{\widehat P}]=i(2\hbar)$.

Although we are interested in the large $\mu$ expansion of the grand potential, the original expression \eqref{Jrhorelation}
\begin{align}
J(\mu)=\sum_{n=1}^\infty\frac{(-1)^{n-1}}{2n}e^{n\mu}{\cal Z}(n),
\label{Jspecial}
\end{align}
with
\begin{align}
{\cal Z}(n)=\tr e^{-n{\widehat H}}
\end{align}
is valid only for small $e^\mu$.
To achieve the large $\mu$ expansion from small $e^\mu$, in \cite{MN2} we utilized a reciprocal formula which follows from
\begin{align}
\sum_{\ell=-\infty}^\infty\frac{(-e^\mu)^\ell}{\ell+\alpha}
=\frac{\pi}{\sin\pi\alpha}e^{-\alpha\mu}.
\end{align}
This manipulation was further generalized in \cite{Ha} by using the Mellin-Barnes integration representation.
Namely, we rewrite the grand potential as an integration
\begin{align}
J(\mu)=-\int^{\epsilon+i\infty}_{\epsilon-i\infty}
\frac{dt}{4\pi i}\Gamma(t)\Gamma(-t){\cal Z}(t)e^{t\mu},
\label{MB}
\end{align}
with $0<\epsilon<1$ and evaluate it in both the regions $\mu>0$ and $\mu<0$.
Assuming $\mu<0$, we can reproduce the series expansion \eqref{Jspecial} by collecting the residues of the integrand in $\mathrm{Re}(t)>\epsilon$.
Assuming $\mu>0$, on the other hand, we can evaluate the integration by pinching the contour so that it encloses the region $\mathrm{Re}(t)<\epsilon$.
As a result, we obtain the large $\mu$ expansion of the grand potential from the residues of the integrand in $\mathrm{Re}(t)<\epsilon$.

To explicitly study the grand potential, we use the WKB $\hbar$-expansion, as in the ABJM theory \cite{MP}.
The $\hbar$-expansion of ${\cal Z}(n)$ takes the form
\begin{align}
{\cal Z}(n)=\frac{1}{\hbar}\sum_{\ell=0}^\infty\hbar^{2\ell}{\cal Z}_\ell(n),
\end{align}
where the overall factor $1/\hbar$ is due to the normalization by the unit volume of the phase space.
Note that, since quantum corrections contain $\hbar$ only through $i\hbar$, the hermiticity of the Hamiltonian ensures that quantum corrections only appear in even powers of $\hbar$.
Correspondingly, we also decompose the grand potential as
\begin{align}
J(\mu)=\frac{1}{\hbar}\sum_{\ell=0}^\infty\hbar^{2\ell}J_\ell(\mu).
\end{align}

Below we first compute the classical limit ${\cal Z}_0(n)$, by neglecting the ordering of the operators and performing the phase space integral explicitly.
Then, using the Mellin-Barnes integration representation \eqref{MB} we obtain the exact large $\mu$ expansion of $J_0(\mu)$.
After that, we proceed to the quantum $\hbar$-corrections and determine $J_2(\mu)$ and $J_4(\mu)$ by the same method.

\subsection{Classical limit}
In the classical limit, all the operators can be regarded as $c$-numbers and the trace is the $(Q,P)$-phase space integral divided by $4\pi\hbar$.
As a result, ${\cal Z}_0(n)$ is
\begin{align}
{\cal Z}_0(n)=\int\frac{dQdP}{4\pi}e^{-nH_{0}},
\label{Z0}
\end{align}
with the classical Hamiltonian $H_0$ given by
\begin{align}
H_{0}=(r-2)U+(r-2)T-2S.
\end{align}
Here $U$, $T$ and $S$ are given by \eqref{UTS} with the operators ${\widehat Q}$ and ${\widehat P}$ replaced simply by $c$-numbers $Q$ and $P$ respectively.
Then, the integration in \eqref{Z0} is found to factorize as
\begin{align}
{\cal Z}_0(n)=\frac{(2n)!}{4\pi}\sum_{a,b\ge 0,a+b=n}
\biggl[\int\frac{dQ}{(2a)!}\frac{\bigl(\sinh\frac{Q}{2}\bigr)^{2a}}
{\bigl(2\cosh\frac{Q}{2}\bigr)^{rn-2b}}\biggr]\cdot
\biggl[\int\frac{dP}{(2b)!}\frac{\bigl(\sinh\frac{P}{2}\bigr)^{2b}}
{\bigl(2\cosh\frac{P}{2}\bigr)^{rn-2a}}\biggr].
\end{align}
Using the integration formula $(a\in\mathbb{Z}_{\ge 0})$
\begin{align}
\int_{-\infty}^\infty\frac{dx}{(2a)!}\frac{\bigl(\sinh\frac{x}{2}\bigr)^{2a}}{\bigl(2\cosh\frac{x}{2}\bigr)^m}
=\frac{\Gamma\bigl(\frac{m}{2}-a\bigr)\Gamma\bigl(\frac{m}{2}\bigr)}{2^{2a}\cdot a!\Gamma(m)},
\end{align}
which can be derived recursively by integration by parts, starting with
\begin{align}
\int_{-\infty}^\infty dx\frac{1}{\bigl(2\cosh\frac{x}{2}\bigr)^m}
=\frac{\Gamma\bigl(\frac{m}{2}\bigr)^2}{\Gamma(m)},
\end{align}
and the formula
\begin{align}
\sum_{a,b\ge 0,a+b=n}
\frac{\Gamma\bigl(\frac{x}{2}-a\bigr)\Gamma(x)}{a!\Gamma\bigl(\frac{x}{2}\bigr)\Gamma(x-2a)}
\frac{\Gamma\bigl(\frac{y}{2}-b\bigr)\Gamma(y)}{b!\Gamma\bigl(\frac{y}{2}\bigr)\Gamma(y-2b)}
=\frac{\Gamma\bigl(\frac{x+y}{2}\bigr)}{2^{-2n}\cdot n!\Gamma\bigl(\frac{x+y}{2}-n\bigr)},
\end{align}
which can be shown by considering the generating function with respect to $n$, we finally obtain the following expression for ${\cal Z}_0(n)$
\begin{align}
{\cal Z}_0(n)=\frac{1}{4\pi}\frac{\Gamma(2n+1)}{\Gamma(n+1)}
\frac{\Gamma\bigl((\frac{r}{2}-1)n\bigr)^2}{\Gamma((r-1)n)}
\frac{\Gamma\bigl(\frac{r}{2}n\bigr)^2}{\Gamma(rn)}.
\end{align}

Plugging this into the Mellin-Barnes representation \eqref{MB} and collecting the residues in $\mathrm{Re}(t)\le 0$, we obtain the exact large $\mu$ expansion of the classical grand potential
\begin{align}
J_0(\mu)=\frac{C_0}{3}\mu^3+B_0\mu+A_0+J_0^\text{np}(\mu).
\end{align}
Here the first three perturbative terms come from the residue at $t=0$.
The coefficients $C_0$ and $B_0$ are consistent with the classical Fermi surface analysis in section \ref{BC}, and the constant $A_0$ is
\begin{align}
A_0=\frac{\zeta(3)}{\pi}\biggl(\frac{1}{r}+\frac{r^2}{r-2}\biggr).
\label{A0}
\end{align}
The non-perturbative part $J_0^\text{np}(\mu)$ consists of three kinds of instantons
\begin{align}
J_0^\text{np}(\mu)=\sum_{\ell=1}^\infty c^{(1)}_\ell e^{-\frac{2\ell\mu}{r}}+\sum_{m=1}^\infty(b^{(2)}_m\mu+c^{(2)}_m)e^{-\frac{2m\mu}{r-2}}+\sum_{n=1}^\infty c^{(3)}_ne^{-\frac{n\mu}{2}},
\end{align}
with
\begin{align}
c^{(1)}_\ell&=-\frac{(2\ell)!}{\pi r(\ell!)^2}\frac{
\Gamma\bigl(\frac{2\ell}{r}\bigr)
\Gamma\bigl(-\frac{4\ell}{r}\bigr)
\Gamma\bigl(-\frac{(r-2)\ell}{r}\bigr)^2
}
{
\Gamma\bigl(-\frac{2(r-1)\ell}{r}\bigr)
},\nonumber\\
b^{(2)}_m&=-\frac{1}{\pi (r-2)^2(m!)^2}
\frac{
\Gamma\bigl(\frac{2m}{r-2}\bigr)
\Gamma\bigl(-\frac{4m}{r-2}\bigr)
\Gamma\bigl(-\frac{rm}{r-2}\bigr)^2
}
{
\Gamma\bigl(-\frac{2(r-1)m}{r-2}\bigr)
\Gamma\bigl(-\frac{2rm}{r-2}\bigr)
},\nonumber\\
\frac{c^{(2)}_m}{b^{(2)}_m}&=
-\psi\Bigl(\frac{2m}{r-2}\Bigr)
+2\psi\Bigr(-\frac{4m}{r-2}\Bigr)
+(r-2)\psi(m+1)
+r\psi\Bigr(-\frac{rm}{r-2}\Bigr)\nonumber\\
&\quad-(r-1)\psi\Bigr(-\frac{2(r-1)m}{r-2}\Bigr)
-r\psi\Bigr(-\frac{2rm}{r-2}\Bigr),
\nonumber\\
c^{(3)}_n&=\frac{(-1)^{n-1}}{8\pi n!}\frac{
\Gamma\bigl(\frac{n}{2}\bigr)
\Gamma\bigl(-\frac{(r-2)n}{4}\bigr)^2
\Gamma\bigl(-\frac{rn}{4}\bigr)^2
}
{
\Gamma\bigl(-\frac{(r-1)n}{2}\bigr)\Gamma\bigl(-\frac{rn}{2}\bigr)
},
\end{align}
where $\psi(x)$ is the di-gamma function $\psi(x)=\partial_x\log\Gamma(x)$.

\subsection{Quantum corrections}
Now we shall go on to the quantum corrections.
As in \cite{MP}, with the help of the Wigner transformation
\begin{align}
({\widehat X})_\text{W}=\int\frac{dQ^\prime}{2\pi}
\biggl\langle Q-\frac{Q^\prime}{2}
\,\biggl|\,{\widehat X}\,\biggr|\,
Q+\frac{Q^\prime}{2}\biggr\rangle
e^{\frac{iQ^\prime P}{2\hbar}},
\label{Wigner}
\end{align}
the trace of operators in ${\cal Z}(n)$ can be expressed as an integration of a $c$-function
\begin{align}
{\cal Z}(n)=\int\frac{dQdP}{4\pi\hbar}(e^{-n{\widehat H}})_\text{W}.
\end{align}
Practically, the Wigner transformation can be computed by
\begin{align}
f({\widehat Q})_\text{W}=f(Q),\quad
f({\widehat P})_\text{W}=f(P),\quad
({\widehat X}{\widehat Y})_\text{W}
=({\widehat X})_\text{W}\star({\widehat Y})_\text{W},
\end{align}
where the star product is given as
$\star=\exp[i\hbar(\overleftarrow{\partial}_Q\overrightarrow{\partial}_P
-\overleftarrow{\partial}_P\overrightarrow{\partial}_Q)]$.
In this formulation, we can compute all the $\hbar$-corrections systematically through the $\star$-product.

As in the case of the ${\widehat A}$-type theories with ${\cal N}=4$ supersymmetry, there are two sources of $\hbar$-corrections: the deviation of $H_\text{W}$ from $H_{0}$, and the deviation of $(e^{-n{\widehat H}})_\text{W}$ from $e^{-nH_\text{W}}$.
The latter can be incorporated in the same way as in the ${\widehat A}$-type theories, by applying the Taylor expansion
\begin{align}
f({\widehat X})_\text{W}=\sum_{\ell=0}^\infty
\frac{1}{\ell!}
\frac{\partial^\ell f(X_W)}{\partial X_W^\ell}
{\cal G}_\ell(X_\text{W}),\quad
{\cal G}_\ell(X_\text{W})
=\bigl(({\widehat X}-X_\text{W})^\ell\bigr)_\text{W},
\label{Wignerformula}
\end{align}
with $f(x)=e^{-nx}$.
The former deviation can also be calculated similarly as in the ${\widehat A}$-type theories, by the Baker-Campbell-Hausdorff formula and the $\star$-product.
In deriving the Hamiltonian operator, all we have to do is to utilize the formula in appendix A of \cite{MP} twice, combining first $e^{\widehat S}$ and then $e^{-\frac{r-2}{2}\widehat U}$ to $e^{-(r-2)\widehat T}$ in \eqref{eHeUeSeT}.
Note that ${\widehat S}$ is a composite of ${\widehat Q}$ and ${\widehat P}$ in this case.
This again can be treated by the formula \eqref{Wignerformula} with $f(x)=\log x$.
In summary, ${\cal Z}(n)$ is given by
\begin{align}
{\cal Z}(n)
&=\int\frac{dQdP}{4\pi\hbar}e^{-nH_{0}} 
\biggl[1+\sum_{\ell=1}^\infty\frac{(-n)^\ell}{\ell!}(H_\text{W}-H_{0})^\ell\biggr]
\biggl[1+\sum_{\ell=2}^\infty\frac{(-n)^\ell}{\ell!}{\cal G}_{\ell}(H_\text{W})\biggr].
\label{calZsystematicexpansion}
\end{align}

Though the calculation is now rather straightforward, let us note that the expression is simplified if we introduce
\begin{align}
U_2=2\cosh\frac{Q}{2},\quad
T_2=2\cosh\frac{P}{2},\quad
S_2=2\sinh\frac{Q+P}{2}.
\end{align}
For example, the $\hbar$-corrections relevant to ${\cal Z}_2(n)$ are given by
\begin{align}
H_\text{W}-H_{0}&=\hbar^2\biggl[
-\frac{(r-2)(r^2+2r+2)}{12U_2^2}
+\frac{(r-2)(r^2+2r+8)}{24T_2^2}
-\frac{(r-2)(r-4)}{12S_2^2}\nonumber \\
&\quad+\frac{(r-2)^2(r+4)}{6U_2^2T_2^2}
-\frac{(r-2)(r+6)}{3U_2^2S_2^2}
+\frac{2r(r-2)}{3T_2^2S_2^2}
-\frac{4U_2^\prime}{3T_2S_2^3}
-\frac{4T_2^\prime}{3U_2S_2^3}\nonumber \\
&\quad-\frac{2(r^2-3r+8)U_2^\prime T_2^\prime}{3U_2T_2S_2^2}
-\frac{2(r-2)(r+2)U_2^\prime S_2^\prime}{3U_2T_2^2S_2}
+\frac{2(r-2)(2r+1)T_2^\prime S_2^\prime}{3U_2^2T_2S_2}
\biggr]+{\cal O}(\hbar^4),\nonumber \\
{\cal G}_2(H_\text{W})&=\hbar^2\biggl[
-\frac{r^2}{U_2^2T_2^2}
-\frac{2r}{U_2^2S_2^2}
-\frac{2r}{T_2^2S_2^2}
\biggr]+{\cal O}(\hbar^4),\nonumber \\
{\cal G}_3(H_\text{W})&=
\hbar^2\biggl[
-\frac{r(r^2+4)}{4U_2^2}
-\frac{r(r^2+4)}{4T_2^2}
-\frac{r^2}{S_2^2}
+\frac{2r^3}{U_2^2T_2^2}
+\frac{2r(r-2)}{U_2^2S_2^2}
+\frac{2r(r-2)}{T_2^2S_2^2}\nonumber \\
&\quad+\frac{4r^2U_2^\prime T_2^\prime}{U_2T_2S_2^2}
+\frac{4r^2U_2^\prime S_2^\prime}{U_2T_2^2S_2}
+\frac{4r^2T_2^\prime S_2^\prime}{U_2^2T_2S_2}
\biggr]+{\cal O}(\hbar^4),
\end{align}
while ${\cal G}_\ell(H_\text{W})={\cal O}(\hbar^4)$ for $\ell\ge 4$.
After substituting these into \eqref{calZsystematicexpansion}, we can integrate the resulting expression by the same technique as in ${\cal Z}_0(n)$.
We finally obtain ${\cal Z}_2(n)$ as
\begin{align}
{\cal Z}_2(n)&=\frac{r^2(r-2)^2n^2(1-n)(1+2n)}{96(1+(r-1)n)(1+rn)}{\cal Z}_0(n).
\end{align}
By a similar, though more lengthy, calculation, we also obtain ${\cal Z}_4(n)$ as
\begin{align}
{\cal Z}_4(n)
&=\frac{r^3(r-2)^2n^3(n-1)(2n+1)}{92160(1+rn)(3+rn)(1+(r-1)n)(2+(r-1)n)(3+(r-1)n)}\nonumber \\
&\hspace{-10mm}\Bigl[(8-5r+r^2)(96+(-110+82r)n+(326-58r+17r^2)n^2+(92+124r-5r^2)n^3+14r^2n^4)\nonumber \\
&\hspace{-5mm}+(-432+226r)n+(-1616+530r)n^2+(-928+144r)n^3-56rn^4\Bigr]{\cal Z}_0(n).
\end{align}

\begin{figure}[h!]
\centering
\includegraphics[width=12cm]{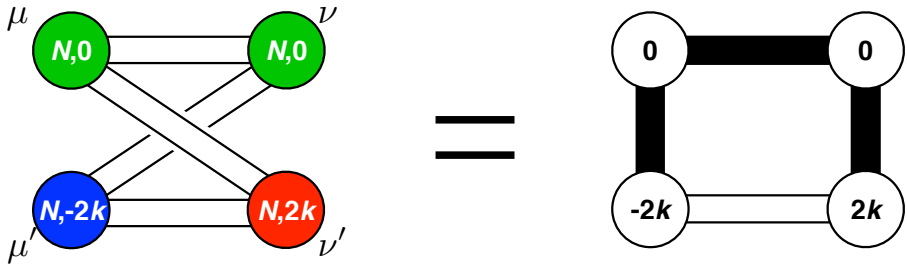}
\caption{Dynkin diagram of $\widehat D_3$.
The edges can be determined e.g.\ from the inner product of the canonical basis
$e_\mu=(1,-1,0), e_{\mu'}=(-1,-1,0), e_\nu=(0,1,-1), e_{\nu'}=(0,1,1)$.
In our parametrization \eqref{levelsforD}, the levels are given by 
$(k_{\nu'},k_\mu,k_\nu,k_{\mu'})=2k(1,0,0,-1)$.
This is nothing but the $\widehat A_3$ Dynkin diagram for the $(3,1)_{2k}$ model, whose array of $s_m$ is given by $(s_{\nu'},s_\mu,s_\nu,s_{\mu'})=(1,1,1,-1)$.
See figure 1 in \cite{MN1}.
}
\label{D3}
\end{figure}

Now let us consider the large $\mu$ expansion of the quantum corrections to the grand potential.
Remarkably, both ${\cal Z}_2(n)$ and ${\cal Z}_4(n)$ are expressed as ${\cal Z}_0(n)$ times some rational function of $n$.
Therefore, $J_2(\mu)$ and $J_4(\mu)$ have the same three species of instantons as $J_0(\mu)$, since the infinite sequences of poles of $\Gamma(t)\Gamma(-t){\cal Z}_0(t)$ in $\mathrm{Re}(t)\le 0$ remain unchanged in the quantum corrections.\footnote{
Though the rational functions have a finite number of poles with $n<0$, all of them are canceled by the zeroes of $[\Gamma(rn)\Gamma((r-1)n)]^{-1}$ in ${\cal Z}_0(n)$ and none produce new instanton effects.
}
On the other hand, the perturbative part which comes from the residue at $n=0$ is
\begin{align}
J_2(\mu)&=\frac{r(r-1)}{48\pi}\mu-\frac{r(r-1)^2}{24\pi}+J_2^{\text{np}}(\mu),\nonumber \\
J_4(\mu)&=-\frac{r^2(r-1)(r^2-5r+8)}{8640\pi}+J_4^{\text{np}}(\mu).
\label{A24}
\end{align}
Combining the classical value \eqref{A0} and the quantum corrections \eqref{A24} for the constant part $A$, we find that the result is consistent with \eqref{Aspecial}, at least up to ${\cal O}(\hbar^5)$.
This is reminiscent of a similar relation discovered for the $(q,p)_k$ models among the theories of the ${\widehat A}$-type quiver in \cite{MN1}.
Note that the Planck constant seems doubled compared with the $(q,p)_k$ models.
This can be understood by comparing two identical quivers $\widehat D_3$ and $\widehat A_3$, where our case with uniform $s_m=1$ is identified with the $(3,1)_{2k}$ model.
See figure \ref{D3}.

\section{Summary and discussion}
In this paper we have studied the partition function of the superconformal Chern-Simons theories of the ${\widehat D}$-type quiver, and have shown that we can rewrite the partition function into that of the Fermi gas system as in the case of the ${\widehat A}$-type quiver.
We find that, again, the perturbative corrections of the partition function are summed up to the Airy function, if the Hamiltonian of the Fermi gas system is hermitian.
Though, for the general $\widehat D$-type quiver, in section \ref{BC} we only consider the perturbative coefficients in the classical limit $k\rightarrow 0$, the Fermi gas formalism is very powerful and allows us in principle to determine the quantum corrections and the non-perturbative instanton corrections.

To further proceed to studying the membrane instanton of the general $\widehat D$-type quivers quantum-mechanically by the WKB expansion, it is, however, difficult to handle the non-commutative operators in the density matrix, or the exact integration over the phase space without taking the large $\mu$ limit.
In the theories of the general ${\widehat A}$-type quivers, we have overcome the difficulties \cite{MN1,MN2,MN3} by restricting ourselves to those with ${\cal N}=4$ supersymmetry \cite{IK4}.
Similarly, here in the theories of the ${\widehat D}$-type quivers, the difficulty is resolved by choosing the quivers with uniform $s_m$, as in section \ref{specialinst}.
For these special values of the levels, we have found that the non-perturbative corrections consist of three kinds of instantons, and have also observed that the constant $A$ can be expressed in terms of that in the ABJM theory (at least up to ${\cal O}(k^5)$).
These are reminiscent of the results for the theories of the ${\widehat A}$-type quivers with the ${\cal N}=4$ supersymmetry \cite{MN1,MN2}.

It is interesting to see whether the symmetry is enhanced for these cases with uniform $s_m$.
Also, we hope to interpret these instanton exponents from the dual supergravity picture, as membranes wrapping on the tri-Sasaki Einstein manifold, though the geometry is more complicated than that for the ${\widehat A}$-type quivers.
Furthermore, we hope to proceed to all the non-perturbative corrections including the worldsheet instantons which have not been discussed at all in this work.

After seeing the Fermi gas formalism for the theories of the $\widehat A$-type and $\widehat D$-type quivers, it should be interesting to ask whether a Fermi gas formalism exists also for the ${\widehat E}$-type quivers.
Also, it is interesting to study other quivers with orthosymplectic groups in \cite{GHN} from the Fermi gas formalism. See e.g.\ \cite{MePu}.

\appendix

\section{A pfaffian formula}\label{detpf}

{\bf Proposition}.
Let $(\phi_a)_{1\le a\le 2N}$ and $(\psi_b)_{1\le b\le 2N}$ be functions on a measurable space.
Then we have
\begin{align}
\int\frac{D^Nx}{N!}
\det\begin{pmatrix}(\phi_a(x_i))
_{\begin{subarray}{c}1\le a\le 2N\\1\le i\le N\end{subarray}}&
(\psi_a(x_i))
_{\begin{subarray}{c}1\le a\le 2N\\1\le i\le N\end{subarray}}
\end{pmatrix}
=(-1)^{\frac{1}{2}(N-1)N}\pf P_{ab},
\label{pfaffian}
\end{align}
with the skew-symmetric matrix $P$
\begin{align}
P_{ab}=\int Dx(\phi_a(x)\psi_b(x)-\phi_b(x)\psi_a(x)).
\end{align}

\noindent
{\bf Remark}.
The definition of the pfaffian for a skew-symmetric matrix $P$ is given by
\begin{align}
\pf P=(-1)^{\frac{1}{2}(N-1)N}\frac{1}{2^NN!}\sum_{\sigma\in S_{2N}}
(-1)^\sigma\prod_{i=1}^NP_{\sigma(i)\sigma(i+N)}.
\end{align}

\noindent
{\bf Proof}.
We can prove it by skew-symmetrizing the matrix elements,
\begin{align}
&\frac{1}{N!}\int D^Nx
\det\begin{pmatrix}(\phi_a(x_i))_{2N\times N}
&(\psi_a(x_i))_{2N\times N}\end{pmatrix}\nonumber\\
&=\frac{1}{N!}\int D^Nx
\sum_{\sigma\in S_{2N}}(-1)^\sigma\prod_{i=1}^N
\phi_{\sigma(i)}(x_i)\psi_{\sigma(i+N)}(x_i)\nonumber\\
&=\frac{1}{N!}
\sum_{\sigma\in S_{2N}}(-1)^\sigma\prod_{i=1}^N\frac{1}{2}
\int Dx\Bigl(\phi_{\sigma(i)}(x)\psi_{\sigma(i+N)}(x)
-\phi_{\sigma(i+N)}(x)\psi_{\sigma(i)}(x)\Bigr)\nonumber\\
&=(-1)^{\frac{1}{2}(N-1)N}\pf P.
\end{align}

\section{Another pfaffian formula}\label{squareroot}
{\bf Proposition}.
Let $P_{ab}(x,y)$ with $a,b=1,2$ be functions of two variables satisfying $P_{ba}(y,x)=-P_{ab}(x,y)$.
Let $\overline P$ be a $2N\times 2N$ skew-symmetric matrix consisting of four $N\times N$ blocks $P_{ab}$ whose $(i,j)$-component is $P_{ab}(x_i,x_j)$.
Then, we have
\begin{align}
\sum_{N=0}^\infty z^N\int\frac{d^Nx}{N!}
(-1)^{\frac{1}{2}(N-1)N}\pf\overline P
=\sqrt{\det\bigl(\overline I-z\,\overline\Omega\,\overline P\bigr)},
\end{align}
with various matrices on the right-hand side defined by the identity operator as
\begin{align}
\overline\Omega=\begin{pmatrix}0&I\\-I&0\end{pmatrix},\quad
\overline I=\begin{pmatrix}I&0\\0&I\end{pmatrix}.
\label{OmegaI}
\end{align}
Here the pfaffian on the left-hand side is the finite dimensional one, while on the right-hand side the determinant denotes simultaneously the $2\times 2$ determinant and the Fredholm determinant.

\noindent
{\bf Remark}.
This is the continuum limit $N_\infty\to\infty$ of the following proposition (See e.g.\ Proposition 2.1 in \cite{M}).
Note that, in taking the limit, we use $\pf(\overline\Omega+z\overline P)^2=\det(\overline\Omega+z\overline P)=\det(\overline I-z\,\overline \Omega\,\overline P)$, which follows from $\overline\Omega^{-1}=-\overline\Omega$ and $\det\overline\Omega=1$, and fix the overall signs by setting $\overline P$ to be zero.

\noindent
{\bf Proposition}.
Let $\overline P_{a,b}$ with $a,b=1,\cdots,2N_\infty$ be a skew-symmetric matrix.
Then, we have
\begin{align}
(-1)^{\frac{1}{2}(N_\infty-1)N_\infty}\pf(\overline\Omega+z\overline P)
=\sum_{N=0}^{N_\infty}z^N(-1)^{\frac{1}{2}(N-1)N}
\sum_{1\le s_1<\cdots<s_N\le N_\infty}\pf\overline{P}_{\{s\}},
\label{prop2disc}
\end{align}
where $\overline P_{\{s\}}$ consists of four $N\times N$ blocks whose $(i,j)$-component is given by $P_{s_i(+N),s_j(+N)}$:
\begin{align}
\overline{P}_{\{s\}}=
\begin{pmatrix}
(P_{s_i,s_j})_{N\times N}&(P_{s_i,s_j+N})_{N\times N}\\
(P_{s_i+N,s_j})_{N\times N}&(P_{s_i+N,s_j+N})_{N\times N}
\end{pmatrix}.
\end{align}

\section*{Acknowledgements}

We are grateful to Sho Matsumoto and Takuya Matsumoto for collaborative discussions at various stages of this work.
We would also like to thank Yasuyuki Hatsuda, Tokiro Numasawa, Masaki Shigemori, Takao Suyama, Kento Watanabe and especially Nadav Drukker for valuable discussions.
The work of S.M.\ is supported by JSPS Grant-in-Aid for Scientific
Research (C) \# 26400245, while the work of T.N.\ is partly supported
by the JSPS Research Fellowships for Young Scientists.


\begin{thebibliography}{99}
\bibitem{W2}
E.~Witten,
``Some comments on string dynamics,''
In *Los Angeles 1995, Future perspectives in string theory* 501-523
[hep-th/9507121].
%
\bibitem{HMRV}
J.~J.~Heckman, D.~R.~Morrison, T.~Rudelius and C.~Vafa,
``Atomic Classification of 6D SCFTs,''
arXiv:1502.05405 [hep-th].
%
\bibitem{ABJM}
O.~Aharony, O.~Bergman, D.~L.~Jafferis and J.~Maldacena,
``N=6 superconformal Chern-Simons-matter theories, M2-branes and their
gravity duals,''
JHEP {\bf 0810}, 091 (2008)
[arXiv:0806.1218 [hep-th]].
%
\bibitem{KWY}
A.~Kapustin, B.~Willett and I.~Yaakov,
``Exact Results for Wilson Loops in Superconformal Chern-Simons
Theories with Matter,''
JHEP {\bf 1003}, 089 (2010)
[arXiv:0909.4559 [hep-th]].
%
\bibitem{GAH} 
D.~R.~Gulotta, J.~P.~Ang and C.~P.~Herzog,
``Matrix Models for Supersymmetric Chern-Simons Theories with an ADE Classification,''
JHEP {\bf 1201}, 132 (2012)
[arXiv:1111.1744 [hep-th]].
%
\bibitem{GW}
D.~Gaiotto and E.~Witten,
``S-Duality of Boundary Conditions In N=4 Super Yang-Mills Theory,''
Adv.\ Theor.\ Math.\ Phys.\ {\bf 13}, 721 (2009)
[arXiv:0807.3720 [hep-th]].
%
\bibitem{HKPT}
C.~P.~Herzog, I.~R.~Klebanov, S.~S.~Pufu and T.~Tesileanu,
``Multi-Matrix Models and Tri-Sasaki Einstein Spaces,''
Phys.\ Rev.\ D {\bf 83}, 046001 (2011)
[arXiv:1011.5487 [hep-th]].
%
\bibitem{GHP}
D.~R.~Gulotta, C.~P.~Herzog and S.~S.~Pufu,
``From Necklace Quivers to the F-theorem, Operator Counting, and
T(U(N)),''
JHEP {\bf 1112}, 077 (2011)
[arXiv:1105.2817 [hep-th]].
%
\bibitem{GHN}
D.~R.~Gulotta, C.~P.~Herzog and T.~Nishioka,
``The ABCDEF's of Matrix Models for Supersymmetric Chern-Simons Theories,''
JHEP {\bf 1204}, 138 (2012)
[arXiv:1201.6360 [hep-th]].
%
\bibitem{CHJ}
P.~M.~Crichigno, C.~P.~Herzog and D.~Jain,
``Free Energy of $D_n$ Quiver Chern-Simons Theories,''
JHEP {\bf 1303}, 039 (2013)
[arXiv:1211.1388 [hep-th]].
%
\bibitem{KT}
I.~R.~Klebanov and A.~A.~Tseytlin,
``Entropy of near extremal black $p$-branes,''
Nucl.\ Phys.\ B {\bf 475}, 164 (1996)
[hep-th/9604089].
%
\bibitem{DMP1}
N.~Drukker, M.~Marino and P.~Putrov,
``From weak to strong coupling in ABJM theory,''
Commun.\ Math.\ Phys.\  {\bf 306}, 511 (2011)
[arXiv:1007.3837 [hep-th]].
%
\bibitem{FHM}
H.~Fuji, S.~Hirano and S.~Moriyama,
``Summing Up All Genus Free Energy of ABJM Matrix Model,''
JHEP {\bf 1108}, 001 (2011)
[arXiv:1106.4631 [hep-th]].
%
\bibitem{KEK}
M.~Hanada, M.~Honda, Y.~Honma, J.~Nishimura, S.~Shiba and Y.~Yoshida,
``Numerical studies of the ABJM theory for arbitrary N at arbitrary
coupling constant,''
JHEP {\bf 1205}, 121 (2012)
[arXiv:1202.5300 [hep-th]].
%
\bibitem{MP}
M.~Marino and P.~Putrov,
``ABJM theory as a Fermi gas,''
J.\ Stat.\ Mech.\  {\bf 1203}, P03001 (2012)
[arXiv:1110.4066 [hep-th]].
%
\bibitem{HMO1}
Y.~Hatsuda, S.~Moriyama and K.~Okuyama,
``Exact Results on the ABJM Fermi Gas,''
JHEP {\bf 1210}, 020 (2012)
[arXiv:1207.4283 [hep-th]].
%
\bibitem{PY}
P.~Putrov and M.~Yamazaki,
``Exact ABJM Partition Function from TBA,''
Mod.\ Phys.\ Lett.\ A {\bf 27}, 1250200 (2012)
[arXiv:1207.5066 [hep-th]].
%
\bibitem{HMO2}
Y.~Hatsuda, S.~Moriyama and K.~Okuyama,
``Instanton Effects in ABJM Theory from Fermi Gas Approach,''
JHEP {\bf 1301}, 158 (2013)
[arXiv:1211.1251 [hep-th]].
%
\bibitem{CM}
F.~Calvo and M.~Marino,
``Membrane instantons from a semiclassical TBA,''
JHEP {\bf 1305}, 006 (2013)
[arXiv:1212.5118 [hep-th]].
%
\bibitem{HMO3}
Y.~Hatsuda, S.~Moriyama and K.~Okuyama,
``Instanton Bound States in ABJM Theory,''
JHEP {\bf 1305}, 054 (2013)
[arXiv:1301.5184 [hep-th]].
%
\bibitem{HMMO}
Y.~Hatsuda, M.~Marino, S.~Moriyama and K.~Okuyama,
``Non-perturbative effects and the refined topological string,''
JHEP {\bf 1409} (2014) 168
[arXiv:1306.1734 [hep-th]].
%
\bibitem{HM}
M.~Honda and S.~Moriyama,
``Instanton Effects in Orbifold ABJM Theory,''
JHEP {\bf 1408} (2014) 091
[arXiv:1404.0676 [hep-th]].
%
\bibitem{MN1}
S.~Moriyama and T.~Nosaka,
``Partition Functions of Superconformal Chern-Simons Theories from Fermi Gas Approach,''
JHEP {\bf 1411} (2014) 164
[arXiv:1407.4268 [hep-th]].
%
\bibitem{MN2}
S.~Moriyama and T.~Nosaka,
``ABJM Membrane Instanton from Pole Cancellation Mechanism,''
arXiv:1410.4918 [hep-th].
%
\bibitem{MN3}
S.~Moriyama and T.~Nosaka,
``Exact Instanton Expansion of Superconformal Chern-Simons Theories from Topological Strings,''
JHEP {\bf 1505} (2015) 022
[arXiv:1412.6243 [hep-th]].
%
\bibitem{MePu}
M.~Mezei and S.~S.~Pufu,
``Three-sphere free energy for classical gauge groups,''
JHEP {\bf 1402}, 037 (2014)
[arXiv:1312.0920 [hep-th], arXiv:1312.0920].
%
\bibitem{GM}
A.~Grassi and M.~Marino,
``M-theoretic matrix models,''
JHEP {\bf 1502} (2015) 115
[arXiv:1403.4276 [hep-th]].
%
\bibitem{HaOk}
Y.~Hatsuda and K.~Okuyama,
``Probing non-perturbative effects in M-theory,''
JHEP {\bf 1410} (2014) 158
[arXiv:1407.3786 [hep-th]].
%
\bibitem{DF} 
N.~Drukker and J.~Felix,
``3d mirror symmetry as a canonical transformation,''
JHEP {\bf 1505} (2015) 004
[arXiv:1501.02268 [hep-th]].
%
\bibitem{DMP2} 
N.~Drukker, M.~Marino and P.~Putrov,
``Nonperturbative aspects of ABJM theory,''
JHEP {\bf 1111}, 141 (2011)
[arXiv:1103.4844 [hep-th]].
%
\bibitem{ADF}
B.~Assel, N.~Drukker and J.~Felix,
``Partition Functions of 3d $\hat D$-Quivers and Their Mirror Duals from 1d Free Fermions,''
arXiv:1504.07636 [hep-th].
%
\bibitem{MM}
S.~Matsumoto and S.~Moriyama,
``ABJ Fractional Brane from ABJM Wilson Loop,''
JHEP {\bf 1403}, 079 (2014)
arXiv:1310.8051 [hep-th].
%
\bibitem{Ha}
Y.~Hatsuda,
``Spectral zeta function and non-perturbative effects in ABJM Fermi-gas,''
arXiv:1503.07883 [hep-th].
%
\bibitem{IK4}
Y.~Imamura and K.~Kimura,
``N=4 Chern-Simons theories with auxiliary vector multiplets,''
JHEP {\bf 0810}, 040 (2008)
[arXiv:0807.2144 [hep-th]].
%
\bibitem{M}
S.~Matsumoto,
``$\alpha$-Pfaffian, pfaffian point process and shifted Schur measure,''
Linear algebra and its applications {\bf 403}, 369 (2005).
%
\end{thebibliography}
\end{document}